\documentclass[conference]{IEEEtran}
\IEEEoverridecommandlockouts
\usepackage{cite}
\usepackage{amsmath,amssymb,amsfonts}
\usepackage{algorithmic}
\usepackage{graphicx}
\usepackage{textcomp}
\usepackage{xcolor}
\usepackage{booktabs, multirow} % for borders and merged ranges
\usepackage{soul}% for underlines
\usepackage{xcolor,colortbl} % for cell colors
\usepackage{changepage,threeparttable} % for wide tables
\usepackage{hyperref}

\definecolor{maroon}{HTML}{F26035}
\definecolor{yellow}{HTML}{FDBC42}
\definecolor{darkred}{RGB}{156, 39, 33}
\definecolor{darkblue}{RGB}{31, 90, 153}
\definecolor{forestgreen}{rgb}{0.13, 0.55, 0.13}
\definecolor{olmoDarkBlue}{HTML}{012e59}
\definecolor{olmoBlue}{HTML}{265ed4}
\definecolor{olmoLightBlue}{HTML}{012e59}
\definecolor{olmoTeal}{HTML}{00d5ff}
\definecolor{olmoYellow}{HTML}{ffbb00}
\definecolor{olmoOrange}{HTML}{ff9100}
\definecolor{maroon}{HTML}{F26035}
\definecolor{yellow}{HTML}{FDBC42}
\definecolor{lavender}{HTML}{734f96}
\definecolor{darkergrey}{HTML}{444444}
\definecolor{midgrey}{HTML}{e6eded}
\definecolor{ai2pink}{HTML}{f0529c}%
\definecolor{ai2midpink}{HTML}{fad3e5}
\definecolor{ai2lightpink}{HTML}{fbecf3}
\definecolor{ai2midwhite}{HTML}{f2e5d9}
\definecolor{ai2offwhite}{HTML}{fbf4ee}
\definecolor{ai2green}{HTML}{0fcb8c}
\definecolor{ai2lightgreen}{HTML}{e7f9f3}
\definecolor{ai2darkgreen}{HTML}{105257}
\definecolor{ai2purple}{HTML}{B932EB}
\definecolor{ai2lightpurple}{HTML}{f7e8fc}
\definecolor{neutralEight}{HTML}{343434}
\definecolor{neutralFive}{HTML}{838383}
\definecolor{neutralThree}{HTML}{bebebe}
\definecolor{neutralOne}{HTML}{dedede}
\definecolor{lightgrey}{HTML}{fafcfc}

\def\BibTeX{{\rm B\kern-.05em{\sc i\kern-.025em b}\kern-.08em
    T\kern-.1667em\lower.7ex\hbox{E}\kern-.125emX}}
\begin{document}

\title{The CMU-AIST submission for the ICME 2025 Audio Encoder Challenge}

\author{
\IEEEauthorblockN{
    Shikhar Bharadwaj\IEEEauthorrefmark{3}$^{*}$\thanks{*Equal contribution.},
    Samuele Cornell\IEEEauthorrefmark{3}$^{*}$,
    Kwanghee Choi\IEEEauthorrefmark{3},
    Hye-jin Shim\IEEEauthorrefmark{3},\\
    Soham Deshmukh\IEEEauthorrefmark{3},
    Satoru Fukayama\IEEEauthorrefmark{2},
    Shinji Watanabe\IEEEauthorrefmark{3}
}
\IEEEauthorblockA{\IEEEauthorrefmark{3}Carnegie Mellon University, USA\\
\IEEEauthorrefmark{2}National Institute of Advanced Industrial Science and Technology (AIST), Japan\\
sbharad2@andrew.cmu.edu}
}

\maketitle

\begin{abstract}
This technical report describes our submission to the ICME 2025 audio encoder challenge.
Our submitted system is built on BEATs, a masked speech token prediction based audio encoder.
We extend the BEATs model using 74,000 hours of data derived from various speech, music, and sound corpora and scale its architecture upto 300 million parameters.
We experiment with speech-heavy and balanced pre-training mixtures to study the impact of different domains on final performance.
Our submitted system consists of an ensemble of the Dasheng 1.2 billion model with two custom scaled-up BEATs models trained on the aforementioned pre-training data mixtures.
We also propose a simple ensembling technique that retains the best capabilities of constituent models and surpasses both the baseline and Dasheng 1.2B.
For open science, we publicly release our trained checkpoints via huggingface\footnote{\url{https://huggingface.co/shikhar7ssu/OpenBEATs-ICME-SOUND} \& \url{https://huggingface.co/shikhar7ssu/OpenBEATs-ICME}}.

\end{abstract}


\section{Introduction}
\label{sec:intro}

Self-supervised learning (SSL) has demonstrated remarkable success in numerous audio processing tasks. Recent models such as BEATs \cite{beats} and Dasheng \cite{dasheng} have pushed the state-of-the-art in tasks such as sound event detection~\cite{cornell2024dcase}, audio scene classification, and audio captioning \cite{dcase24aac_top}.
In this challenge submission, we experiment with the BEATs model by expanding its pre-training corpus to include multiple datasets totaling 74,000 hours of audio spanning general audio, speech-centered, and music-centered data. We investigated the impact of data composition on model performance by comparing speech-heavy versus balanced sampling strategies from this comprehensive data pool. Our experiments with model ensembling reveal that simple upsampling and concatenation of embedding vectors effectively preserves the strengths of all constituent models. Our final system implements this ensembling approach, combining two distinct scaled-up \texttt{large} BEATs models, each trained on different data distributions (speech-heavy and balanced), with the Dasheng 1.2B model.

\section{System Description}

\subsection{Background}
Our system is based on the BEATs (Bidirectional Encoder representation from Audio Transformers)~\cite{beats} and Dasheng \cite{dasheng} frameworks.
BEATs is a self-supervised audio representation learning framework which adopts the bidirectional masked language model (MLM) training approach pioneered by BERT \cite{bert} in natural language proessing, motivated by the cloze task \cite{cloze}.
BEATs training is based on two-stage self-supervision, which involves teacher encoder and student tokenizer training.
Both teacher and student iteratively refine each other, improving both real-valued representations and discrete token representations.
The backbone of the model is transformer-based and leverages a Vision Transformer architecture (ViT)~\cite{vit}.
Our training code is based on OpenBEATs \cite{bharadwaj2025openbeats}, the open-source implementation of BEATs, while we use the X-ARES toolkit \cite{xares} for evaluations.

\begin{table*}[!htp]\centering
\caption{MLP track results. We used X-ARES toolkit \cite{xares} for evaluations. Challenge Baseline denotes the best challenge baseline among Dasheng-Base, Data2Vec and Whisper.}
\label{tab:mlp}
% \scriptsize
\begin{tabular}{lcccccccc}\toprule
\textbf{Task} & \textbf{Domain} & \textbf{Challenge Baseline} & \textbf{BEATs (90M) } &\textbf{Dasheng} & \textbf{BEATs (300M)} & \textbf{BEATs (300M)} & \textbf{Ensemble} \\

 & & & \textbf{iter3} \cite{beats} & \textbf{1.2B} \cite{dasheng} & \textbf{Balanced} & \textbf{Speech} & \textbf{(Submission)} \\
\midrule
\rowcolor{ai2offwhite} FSD50k &Sound & 0.408 & 0.217 & 0.455 &0.380 &0.432 &\textbf{0.463} \\
Vocal Imitation & Sound &0.238 & 0.212 &0.293 &0.214 &0.223 &\textbf{0.295} \\
 \rowcolor{ai2offwhite}FSD18-Kaggle & Sound &0.557 & 0.545 & 0.627 &0.689 &0.612 &\textbf{0.764} \\
 DESED &Sound & 0.532 & 0.560 & 0.563 &0.552 &0.551 &\textbf{0.566} \\
\rowcolor{ai2offwhite} ESC-50 &Sound & 0.869 & 0.835 & 0.891 &0.868 &0.857 &\textbf{0.904} \\
Clotho &Sound & 0.033 & 0.042 & 0.036 &0.040 &\textbf{0.041} &0.038 \\
\rowcolor{ai2offwhite} UrbanSound 8k & Sound &0.835 & 0.853 & 0.846 &\textbf{0.863} &0.857 &0.862 \\
\midrule
\rowcolor{ai2lightgreen} NSynth-Instruments &Music & 0.693 & 0.579 &0.660 &0.589 &0.550 &\textbf{0.729} \\
 GTZAN Genre &Music & 0.869 & 0.836 & 0.886 &0.859 &0.845 &\textbf{0.898} \\
\rowcolor{ai2lightgreen} Free Music Archive Small & Music & 0.640 & 0.614 &\textbf{0.647} &0.624 &0.616 &0.637 \\
\midrule
 \rowcolor{ai2lightpurple} LibriCount &Speech & 0.688 & 0.665 & 0.728 &0.699 &0.705 &\textbf{0.747} \\
CREMA-D &Speech & 0.772 & 0.642 & 0.790 &0.659 &0.670 &\textbf{0.815} \\
  \rowcolor{ai2lightpurple}RAVDESS &Speech & 0.725 & 0.564 & \textbf{0.793} &0.630 &0.655 &0.792 \\
 Fluent Speech Commands & Speech & 0.962 & 0.545 &\textbf{0.973} &0.585 &0.700 &0.956 \\
 \rowcolor{ai2lightpurple} LibriSpeech-MF &Speech & 0.985 & 0.970 & 0.975 &0.973 &\textbf{0.986} &0.985 \\
Speech Commands V1 &Speech & 0.967 & 0.910 &\textbf{0.973} &0.944 &0.958 &0.972 \\
 \rowcolor{ai2lightpurple} VoxLingua33 &Speech & 0.855 & 0.398 &\textbf{0.860} &0.480 &0.615 &0.817 \\
 VocalSound &Speech & 0.910 & 0.865 &\textbf{0.925} &0.877 &0.879 &0.909 \\
\bottomrule
\end{tabular}
\end{table*}

\begin{figure}[t]
\includegraphics[width=6.5cm]{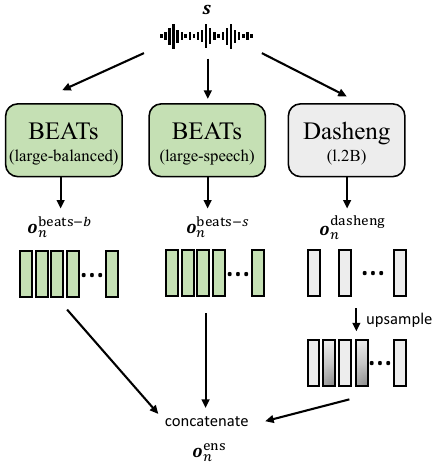}\centering
\caption{Block diagram of the submitted system: we concatenate the output embedding feature vectors from three different systems. Two are based on the BEATs framework and one is the pre-trained Dasheng 1.2B model. We upsample the Dasheng embedding sequence to match the count of embeddings from the BEATs models and concatenate the outputs of the three models. The BEATs models are scaled up to 300M parameter and trained on different pre-training mixtures (speech heavy and balanced) from a 74\,k hour data pool.\label{fig:ensemble}}
\end{figure}

\section{Method}

For our challenge submission, we ensemble three models. The first is Dasheng 1.2B~\cite{dasheng}, selected for its exceptional performance on the X-ARES benchmark~\cite{xares}. We complement this with two custom models based on the BEATs MLM token prediction framework, each trained on a distinct data composition drawn from our multi-domain pre-training corpus. These models represent scaled-up variants we implemented ourselves with 300\,M parameters, compared to the 90\,M parameters in the original proposed BEATs \texttt{base} model~\cite{beats}. This parameter increase is achieved by implementing a 300\,M ViT-Large inspired architecture for the encoder component. To accelerate training for these \texttt{large} BEATs models, we bootstrap the process using a \texttt{base} BEATs iteration 3 tokenizer that we trained exclusively on AudioSet.

\subsection{Ensembling Technique}

Our ensembling technique is illustrated in Figure~\ref{fig:ensemble}.
It addresses two key challenges: differences in output sequence rates and embedding dimension. 
Different audio encoder models, due to differences in architectures or features used (e.g. the size of the patch), may output embeddings  at different frame rates, resulting in a mismatch in the output sequence length. More formally, we can denote with $\{\mathbf{o}_n \in \mathbb{R}^{h}\}_{n=1}^N$ the $h$-dimensional output embedding sequence of length $N$ outputted by an audio encoder model. 
In practice, for different models, we can have mismatched $h$ and $N$ dimensions with the same input signal. 
% For example, Dasheng 1.2B outputs an embedding vector for each 50\,ms, while BEATs outputs a vector for each 20\,ms of input audio, resulting thus in a shorter output sequence for the same input audio length.
To resolve such output sequence length mismatch, as illustrated in Figure~\ref{fig:ensemble}, we upsample the Dasheng embeddings $\mathbf{o}_n^\text{dasheng}$ to match the count with that from the two BEATs models $\mathbf{o}_n^\text{beats-b}$ and  $\mathbf{o}_n^\text{beats-s}$.

Afterwards, we then concatenate the three embedding sequences along the feature axis, obtaining a representation which is effectively the combination of the one from all the models in the ensemble. 
This simple approach proves highly effective. Most downstream applications incorporate a fully connected layer as their initial processing block, whether for linear probing or when training a Transformer on top. In this context, concatenation allows the downstream network to selectively attend to the most relevant features from each model, effectively disregarding less informative representations when necessary. Although we experimented with alternative approaches such as averaging, concatenation consistently delivered superior results.

\subsection{Pre-training Datasets}

For pre-training, we curated a diverse collection of audio datasets totaling 74\,k hours. 
The data is described in Table~\ref{tab:ptdata}.
Our data pool consists of 10\,k hours of general audio sounds, 12\,k hours of music and 52\,k hours of speech-centered data covering both multilingual and English corpora.
All sources were carefully filtered to comply with the challenge constraints, ensuring that no evaluation data was included in the pre-training corpus.
We do not do any data augmentation during pre-training.
From this pool, we constructed two distinct mixtures. 
The primary difference between them is the inclusion of a 34\,k hour subset of YODAS speech data~\cite{yodas}.
We randomly sample the YODAS multilingual training set from OWSM \cite{owsm, owsmv31, owsmctc} to derive this subset.
By including this YODAS data, we obtain a speech-heavy mixture with a 70:15:15 ratio across speech, music, and sound, while excluding it yields a more balanced 40:30:30 distribution.
This ablation enables us to analyze the influence of speech-dominant data on audio representation learning as well as obtaining a speech specialized model and another which is more specialized towards general sound.

\begin{table}[t]\centering
\caption{Training data statistics}\label{tab:ptdata}
% \scriptsize
\begin{tabular}{lrrr}\toprule
\textbf{Datasource} &\textbf{Domain} &\textbf{Hours} \\\midrule
AudioSet \cite{audioset} &Sound &5,000 \\
Freesound &Sound &4,648 \\
BBC Soundeffects &Sound &1,000 \\
VGGSound \cite{vggsound} &Sound &548 \\
Cochlscene \cite{cochlscene} &Sound &169 \\
EpicKitchen \cite{epickitchen} &Sound &157 \\
\midrule
FMA \cite{fma} &Music &7,824 \\
MTG-Jamendo \cite{mtg} &Music &3,701 \\
\midrule
YODAS \cite{yodas} &Speech &34,759 \\
Commonvoice \cite{commonvoice} &Speech &16,304 \\
EARS \cite{ears} &Speech &77 \\
\midrule
\textbf{Total} & & \textbf{74,187} \\
\bottomrule
\end{tabular}
\end{table}

\section{Results}

First, we observe that BEATs scales well with more data and parameters beating the original 90\,M \texttt{base} BEATs model on almost all tasks.
Dasheng 1.2B outperforms challenge baselines by a significant margin, establishing a strong reference point. 
Among our BEATs-style models, the one trained on the speech-heavy mixture, as expected, demonstrates superior performance on speech-related tasks.
This demonstrates the importance of data sampling and ratio in pre-training mixture for achieving good performance towards a particular domain, even when the model is scaled in size. 
Notably, our ensemble model successfully integrates the strengths of all individual models, often matching or surpassing their performance on various tasks, thus validating the effectiveness of our ensembling strategy.

\section{Conclusion}

In our submission, we scaled the pre-training corpus of BEATs to 74,000 hours of data, analyzed the effects of diverse pre-training data mixtures with varying ratios across speech, music, and general sound, and developed a simple yet effective ensembling method.
Our findings suggest that the performance of SSL-based audio encoders is significantly influenced by both the domains represented in pre-training data and their relative proportions.
As such, our final submission
is an ensemble that combines two BEATs models (trained on different data mixtures) with Dasheng 1.2B through upsampling and feature concatenation. This approach effectively leverages the complementary strengths of domain-specialized models. 
 Our results indicate that a possible promising research direction is the use of mixture-of-experts approaches in the audio encoder architecture instead of using ensembles.

\section*{Acknowledgment}
This study was supported by the BRIDGE program of the Cabinet Office, Government of Japan.
Also, we used the Bridges2 system at PSC and Delta system at NCSA through allocation CIS210014 from the Advanced Cyberinfrastructure Coordination Ecosystem: Services \& Support (ACCESS) program, which is supported by National Science Foundation grants \#2138259, \#2138286, \#2138307, \#2137603, and \#2138296.

\bibliographystyle{IEEEbib}
\bibliography{icme2025references}

@inproceedings{bert,
    title = "{BERT}: Pre-training of Deep Bidirectional Transformers for Language Understanding",
    author = "Devlin, Jacob  and
      Chang, Ming-Wei  and
      Lee, Kenton  and
      Toutanova, Kristina",
    editor = "Burstein, Jill  and
      Doran, Christy  and
      Solorio, Thamar",
    booktitle = "Proceedings of the 2019 Conference of the North {A}merican Chapter of the Association for Computational Linguistics: Human Language Technologies, Volume 1 (Long and Short Papers)",
    month = jun,
    year = "2019",
    address = "Minneapolis, Minnesota",
    publisher = "Association for Computational Linguistics",
    url = "https://aclanthology.org/N19-1423/",
    doi = "10.18653/v1/N19-1423",
    pages = "4171--4186",
}

@article{cloze,
  title={“Cloze procedure”: A new tool for measuring readability},
  author={Taylor, Wilson L},
  journal={Journalism quarterly},
  volume={30},
  number={4},
  pages={415--433},
  year={1953},
  publisher={SAGE Publications Sage CA: Los Angeles, CA}
}

@inproceedings{beats,
  title={BEATs: audio pre-training with acoustic tokenizers},
  author={Chen, Sanyuan and Wu, Yu and Wang, Chengyi and Liu, Shujie and Tompkins, Daniel and Chen, Zhuo and Che, Wanxiang and Yu, Xiangzhan and Wei, Furu},
  booktitle={Proceedings of the 40th International Conference on Machine Learning},
  pages={5178--5193},
  year={2023}
}

@inproceedings{owsmctc,
    title = "{OWSM}-{CTC}: An Open Encoder-Only Speech Foundation Model for Speech Recognition, Translation, and Language Identification",
    author = "Peng, Yifan  and
      Sudo, Yui  and
      Shakeel, Muhammad  and
      Watanabe, Shinji",
    editor = "Ku, Lun-Wei  and
      Martins, Andre  and
      Srikumar, Vivek",
    booktitle = "Proceedings of the 62nd Annual Meeting of the Association for Computational Linguistics (Volume 1: Long Papers)",
    month = aug,
    year = "2024",
    address = "Bangkok, Thailand",
    publisher = "Association for Computational Linguistics",
    url = "https://aclanthology.org/2024.acl-long.549/",
    doi = "10.18653/v1/2024.acl-long.549",
    pages = "10192--10209"
}

@INPROCEEDINGS{owsm,
  author={Peng, Yifan and Tian, Jinchuan and Yan, Brian and Berrebbi, Dan and Chang, Xuankai and Li, Xinjian and Shi, Jiatong and Arora, Siddhant and Chen, William and Sharma, Roshan and Zhang, Wangyou and Sudo, Yui and Shakeel, Muhammad and Jung, Jee-Weon and Maiti, Soumi and Watanabe, Shinji},
  booktitle={2023 IEEE Automatic Speech Recognition and Understanding Workshop (ASRU)}, 
  title={Reproducing Whisper-Style Training Using An Open-Source Toolkit And Publicly Available Data}, 
  year={2023},
  volume={},
  number={},
  pages={1-8},
  doi={10.1109/ASRU57964.2023.10389676}}

@inproceedings{audioset,
  title={Audio set: An ontology and human-labeled dataset for audio events},
  author={Gemmeke, Jort F and Ellis, Daniel PW and Freedman, Dylan and Jansen, Aren and Lawrence, Wade and Moore, R Channing and Plakal, Manoj and Ritter, Marvin},
  booktitle={2017 IEEE international conference on acoustics, speech and signal processing (ICASSP)},
  pages={776--780},
  year={2017},
  organization={IEEE}
}

@inproceedings{cochlscene,
  title={CochlScene: Acquisition of acoustic scene data using crowdsourcing},
  author={Jeong, Il-Young and Park, Jeongsoo},
  booktitle={2022 Asia-Pacific Signal and Information Processing Association Annual Summit and Conference (APSIPA ASC)},
  pages={17--21},
  year={2022},
  organization={IEEE}
}

@article{cornell2024dcase,
  title={{DCASE} 2024 task 4: Sound event detection with heterogeneous data and missing labels},
  author={Cornell, Samuele and Ebbers, Janek and Douwes, Constance and Mart{\'\i}n-Morat{\'o}, Irene and Harju, Manu and Mesaros, Annamaria and Serizel, Romain},
  journal={DCASE Workshop},
  year={2024}
}

@techreport{dcase24aac_top,
    author = "Jung, Jee-weon and Zhang, Dong and Yang, Huck C.-H. and Wu, Shih-Lun and Chan, David M. and Kong, Zhifeng and Ruifan, Deng and Yaqian, Zhou and Rafael, Valle and Watanabe, Shinji",
    title = "AUTOMATIC AUDIO CAPTIONING WITH ENCODER FUSION, MULTI-LAYER AGGREGATION, AND LARGE LANGUAGE MODEL ENRICHED SUMMARIZATION",
    institution = "DCASE2024 Challenge",
    year = {2024}
}

@inproceedings{owsmv31,
  title     = {OWSM v3.1: Better and Faster Open Whisper-Style Speech Models based on E-Branchformer},
  author    = {Yifan Peng and Jinchuan Tian and William Chen and Siddhant Arora and Brian Yan and Yui Sudo and Muhammad Shakeel and Kwanghee Choi and Jiatong Shi and Xuankai Chang and Jee-weon Jung and Shinji Watanabe},
  year      = {2024},
  booktitle = {Interspeech 2024},
  pages     = {352--356},
  doi       = {10.21437/Interspeech.2024-1194},
  issn      = {2958-1796},
}

@article{vit,
  title={An image is worth 16x16 words: Transformers for image recognition at scale},
  author={Alexey, Dosovitskiy},
  journal={ICLR},
  year={2020}
}

@article{epickitchen,
  title={The epic-kitchens dataset: Collection, challenges and baselines},
  author={Damen, Dima and Doughty, Hazel and others},
  journal={IEEE Transactions on Pattern Analysis and Machine Intelligence},
  volume={43},
  number={11},
  pages={4125--4141},
  year={2020},
  publisher={IEEE}
}

@inproceedings{fma,
  title={FMA: A Dataset For Music Analysis},
  author={Defferrard, Micha{\"e}l and Benzi, Kirell and Vandergheynst, Pierre and Bresson, Xavier},
  booktitle={International Society for Music Information Retrieval Conference},
  year={2017}
}

@article{dasheng,
  title={Scaling up masked audio encoder learning for general audio classification},
  author={Dinkel, Heinrich and Yan, Zhiyong and Wang, Yongqing and Zhang, Junbo and Wang, Yujun and Wang, Bin},
  journal={Interspeech},
  year={2024}
}

@misc{xares,
  author       = {Junbo Zhang and others},
  title        = {{X-ARES}: eXtensive Audio Representation and Evaluation Suite},
  year         = {2025},
  howpublished = {\url{https://github.com/jimbozhang/xares}},
}

@inproceedings{vggsound,
  title={Vggsound: A large-scale audio-visual dataset},
  author={Chen, Honglie and Xie, Weidi and Vedaldi, Andrea and Zisserman, Andrew},
  booktitle={ICASSP 2020-2020 IEEE International Conference on Acoustics, Speech and Signal Processing (ICASSP)},
  pages={721--725},
  year={2020},
  organization={IEEE}
}

@article{mtg,
  title={The MTG-Jamendo Dataset for Automatic Music Tagging},
  author={Bogdanov, Dmitry and Won, Minz and Tovstogan, Philip and Porter, Alastair and Serra, Xavier},
  year={2019},
  journal={ICML},
}

@inproceedings{yodas,
  title={Yodas: Youtube-oriented dataset for audio and speech},
  author={Li, Xinjian and Takamichi, Shinnosuke and Saeki, Takaaki and Chen, William and Shiota, Sayaka and Watanabe, Shinji},
  booktitle={2023 IEEE Automatic Speech Recognition and Understanding Workshop (ASRU)},
  pages={1--8},
  year={2023},
  organization={IEEE}
}

@inproceedings{commonvoice,
  title={Common Voice: A Massively-Multilingual Speech Corpus},
  author={Ardila, Rosana and Branson, Megan and Davis, Kelly and Kohler, Michael and Meyer, Josh and Henretty, Michael and Morais, Reuben and Saunders, Lindsay and Tyers, Francis and Weber, Gregor},
  booktitle={Proceedings of the Twelfth Language Resources and Evaluation Conference},
  pages={4218--4222},
  year={2020}
}

@article{ears,
  title={EARS: An anechoic fullband speech dataset benchmarked for speech enhancement and dereverberation},
  author={Richter, Julius and Wu, Yi-Chiao and Krenn, Steven and Welker, Simon and Lay, Bunlong and Watanabe, Shinji and Richard, Alexander and Gerkmann, Timo},
  journal={arXiv preprint arXiv:2406.06185},
  year={2024}
}

@article{bharadwaj2025openbeats,
  title={OpenBEATs: A Fully Open-Source General-Purpose Audio Encoder},
  author={Bharadwaj, Shikhar and Cornell, Samuele and Choi, Kwanghee and Fukayama, Satoru and Shim, Hye-jin and Deshmukh, Soham and Watanabe, Shinji},
  journal={arXiv preprint arXiv:2507.14129},
  year={2025}
}

\end{document}